\documentclass[aps,pra,twocolumn]{revtex4}
\usepackage{latexsym}
\usepackage{amsmath}
\usepackage{graphicx}

\newcommand{\be}{\begin{equation}}
\newcommand{\ee}{\end{equation}}

\begin{document}

\title{BEC in a star-comb graph}

\author{F. P. Mancini}
 \author{P. Sodano}
 \author{A. Trombettoni}
\affiliation{
Dipartimento di Fisica and
 Sezione I.N.F.N., Universit\`a di
Perugia, Via A. Pascoli, Perugia, 06123, Italy}

\begin{abstract}
We investigate the properties of free bosons hopping on a
star-comb network, discussing the single-particle spectrum and the
main thermodynamic equilibrium properties: Bose-Einstein critical
temperature, fraction of condensate, and spatial boson
distribution. We find an enhancement of the critical temperature
with respect to other inhomogeneous networks.
\end{abstract}


\maketitle

Although it is well known that free bosons hopping on
translationally invariant networks cannot undergo Bose-Einstein
condensation at finite temperature if the space dimension $d$ is
less or equal to two, very recent studies
\cite{burioni00,burioni01,giusiano04,brunelli04} hint to the
exciting possibility  that the network topology may act as a
catalyst for inducing a finite temperature spatial Bose-Einstein
condensation even if $d<2$. This is indeed possible if one resorts
to a suitable discrete inhomogeneous ambient space on which bosons
are defined. As an example of this situation we shall investigate
the properties of non-interacting bosons hopping on a star-comb
shaped network, evidencing that also in this case one may have a
macroscopic occupation of the ground-state at low temperatures.
Ultracold bosons in optical networks represent a physical system
where this setup could be experimentally implemented.


The star-comb graph is a bundled graph which can be obtained by
grafting a star graph to each site of a linear chain, called
backbone (see Fig. \ref{fig1}). The sites of the graph can be
naturally labelled introducing three integer indices $(x,y,z)$
with $x,y,z\in Z$, where $x=1,\dots, p$ labels the different arms
on each star (excluding the links connecting the different stars),
$y=0,\dots,L$ represents the distance from the backbone  and
$z=1,\dots,N_s$ labels the different stars. In the following we
shall assume periodic boundary conditions
$(x,y,1)\equiv(x,y,N_s+1)$. The Hamiltonian describing
non-interacting bosons hopping on a star-comb graph can be written
as:
\begin{equation}
H=-t\sum_{x,y,z;\:x',y',z'} A_{x,y,z;\:x',y',z'}\:
\hat{a}^{\dag}_{x,y,z} \hat{a}_{x',y',z'} \; .
\label{pure_hopping}
\end{equation}
In Eq. \eqref{pure_hopping}, $t$ is the hopping parameter while
$\hat{a}_{x,y,z}^{\dag}$ ($\hat{a}_{x,y,z}$) is the creation
(annihilation) operator for bosons;  $\hat{n}_{x,y,z}=
\hat{a}_{x,y,z}^{\dag}\hat{a}_{x,y,z}$ is the number operator at
site $(x,y,z)$. The filling, i.e., the average number of particles
per site, is defined as $f=N_T/N_s(p L+1)$, where $N_T$ is the
total number of bosons and $N_s(p L+1)$ is the number of sites.
The adjacency matrix of a star-comb graph is given by:
\begin{equation}
\begin{split}
&A_{x,y,z;\:x',y',z'}= (\delta_{y',y-1}+\delta_{y',y+1})
\;(1-\delta_{y,0}) \delta_{x,x'} \delta_{z,z'}
\\
&+ \delta_{y,0}\; \delta_{y',1} \delta_{z,z'} + \delta_{y,0}\;
\delta_{y',0} \delta_{x,x'} (\delta_{z',z-1}+\delta_{z',z+1}).
\end{split}
\label{adjacency}
\end{equation}
The eigenvalue equation $-t \sum_{j} A_{ij}\psi(j)=E \psi(i)$,
with the previous labelling of sites, reads:
\begin{equation}
\begin{split}
&-t\sum_{x',y',z'}\big[(\delta_{y',y-1}+\delta_{y',y+1})
\;(1-\delta_{y,0}) \delta_{x,x'} \delta_{z,z'}\\
&+ \delta_{y,0}\; \delta_{y',1} \delta_{z,z'} + \delta_{y,0}\;
\delta_{y',0} \delta_{x,x'} (\delta_{z',z-1}+\delta_{z',z+1})\big]
\\
&\cdot \psi(x',y',z')=E \psi(x,y,z)
\end{split}
\label{eq1b}
\end{equation}
By exploiting the translation invariance in the direction of the
backbone, a Fourier transform in the variable $z$ reduces Eq.
(\ref{eq1b}) to a 1-dimensional eigenvalue problem. If one
defines: $\psi(x,y,k)=\sum_{z} e^{ikz}\psi(x,y,z)$, with $k=2\pi
n/N_s$ and $n=1,\dots,N_s$, the eigenvalues equation (\ref{eq1b})
becomes:
\begin{figure}[t]
\includegraphics[scale=0.3]{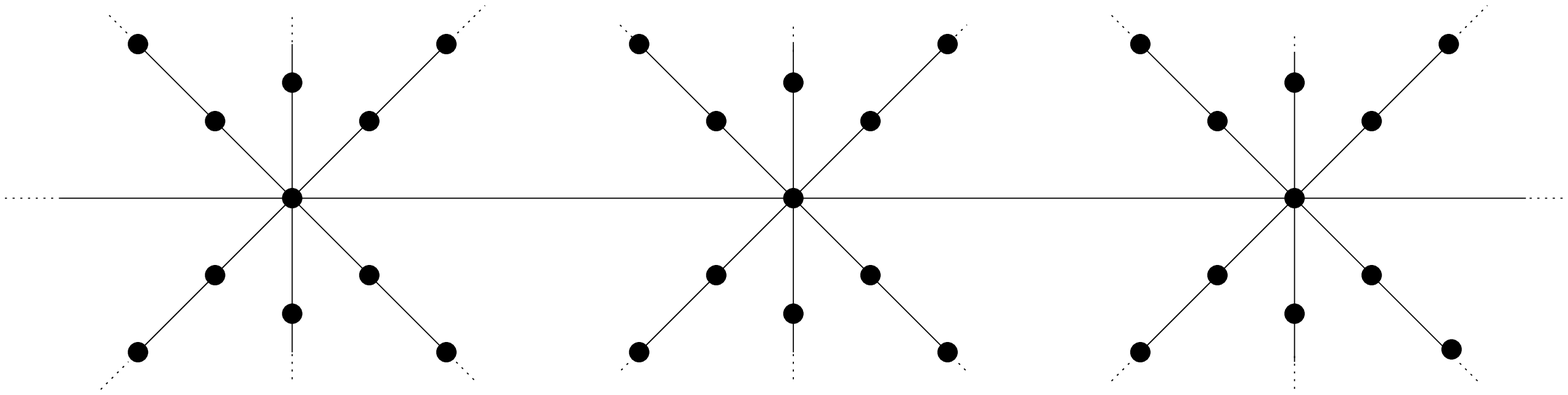}
\caption{\label{fig1} The star-comb graph.}
 \vspace{8mm}
\end{figure}
\begin{equation}
\begin{split}
&-t\sum_{x',y'}\big[ (\delta_{y',y-1}+\delta_{y',y+1})
\;(1-\delta_{y,0}) \delta_{x,x'} + \delta_{y,0}\; \delta_{y',1}
\\
&+2 \cos(k_0) \delta_{y,0}\; \delta_{y',0} \delta_{x,x'} \big]
\psi(x',y')=E \psi(x,y)
\end{split}
\label{eq3b}
\end{equation}
where $\psi(x,y,k)=\delta(k-k_0)\psi(x,y)$, with $k_0=2\pi n/N_s$,
$n=0\dots N_s$. Equation (\ref{eq3b}) can be regarded as the
equivalent problem of a quantum particle hopping on a single star
and interacting with a potential in the center, $V(k_0)=-2t
\cos(k_0)$. The eigenvalues and eigenvectors of Eq. (\ref{eq3b})
are determined by requiring that the free particle solutions on
the arms are solutions also in $y=0$. Indeed, for $y\not =0$, the
wave function satisfying Eq. \eqref{eq3b} is:
\begin{eqnarray}
\psi(x,y)&=&A_x e^{ihy}+B_x e^{-ihy}
 \label{sol1}
\end{eqnarray}
Requiring that - on each arm - $\psi(x,y)$ is a solution of the
eigenvalue equation at $y=L$ introduces restrictions on the
parameters $h$, $A_x$, $B_x$, and on the eigenvalues $E$. In other
words, it amounts to require that $A_x$, $B_x$ and $h$ should
satisfy the $p$ equations
\begin{equation}
\label{hopping_bordo}
\begin{split}
 &-t\left(A_x e^{ih(L-1)}+B_x
e^{-ih(L-1)}\right) \\
&=-2t\cos(h)\left (A_x e^{ihL}+B_x e^{-ihL}\right)
\\
&\quad \quad \quad\quad \quad\quad\quad \quad\quad\quad\quad
\quad\quad \quad \quad \quad x=1,\cdots,p.
\end{split}
\end{equation}
The eigenstates defined in Eq. (\ref{sol1}) should also satisfy
$(p-1)$ matching conditions in the center of each star. Thus, one
has
\begin{equation}
\label{matching} A_x+B_x=A_{x+1}+B_{x+1} \qquad x=1,\cdots,p-1.
\end{equation}
Using Eqs. \eqref{matching}, the condition in one of the centers
gives one more equation
\begin{equation}
\label{hopping_centro}
\begin{split}
&- t\left[ \sum_{x=1}^{p}\left(A_x e^{ih}+B_xe^{-ih}\right) +2
\cos(k_0)\left(A_{x}+B_{x}\right) \right] \\
&= -2t\cos(h)\left(A_{x'}+B_{x'}\right),
\end{split}
\end{equation}
with $x=1,\cdots,p $. Equations (\ref{hopping_bordo}) and
(\ref{hopping_centro}) may be grouped in a homogeneous linear
system of $2 p$ equations which allows to fix the $2p$ parameters
$A_x$ and $B_x$. Upon denoting with $Q$ the ($2p \times 2p$)
matrix whose elements are the coefficients of the linear system
given by Eqs. (\ref{hopping_bordo}) and (\ref{hopping_centro}),
and requiring that
\begin{equation}
\begin{split}
\det Q&=\Theta(h,L) (1-e^{2ih(L+1)})^{p-1} \cdot \big\{
(p-2)\cos(h) \\
&-p  \cot{[h (L+1)]} \sin{(h)}+2 \cos(k_0) \big\} \label{eq_k}
\end{split}
\end{equation}
is equal to zero, the solution is unique. In Eq. (\ref{eq_k})
$\vert \Theta(h,L) \vert=1$ for any value of $h$.

One immediately sees that the values of $h$ for which
$h=n\pi/(L+1)$ (with $n=1,2,...,L$) provide a set of $L
(p-1)$-fold degenerate eigenstates of Eq. (\ref{eq_k}) for each
value of $k_0$ (in total $N_s$ sets). In addition, the solutions
of the transcendental equation
\begin{equation}
(p-2)\cos(h)-p  \cot{[h (L+1)]} \sin{(h)}+2 \cos(k_0)=0
\label{non_degenere}
\end{equation}
provide the values of $h$ associated to non-degenerate
eigenstates. Equation (\ref{non_degenere}) can be solved
numerically and yields a set of  $N_s(L-1)$ non-degenerate
eigenvalues corresponding to different values of $k_0$ and  to
values of $h$ which - in the thermodynamic limit - are equally
spaced and separated by a distance $\pi/(L+1)$. As a result, each
set of delocalized states is formed by $(pL-1)$ states
corresponding to energies ranging between $-2t$ and $+2t$. If the
potential in the center $V(k_0)$ is equal to zero, one recovers
the same condition of the single star \cite{brunelli04}.

Since the total number of states should equal $N_s (p L+1)$, there
are still $2N_s$ localized states in the spectrum. These states
belong to the so-called hidden spectrum \cite{burioni00}. To find
them, it is convenient to look for solutions of the eigenvalue
equation (\ref{eq1b}) of the form
\begin{equation}
\begin{split}
\psi_- (y) &= A e^{-\eta y}+ B e^{\eta y}
\\
\psi_+ (y)&= A (-1)^y e^{-\eta y}+ B (-1)^y  e^{\eta y} ,
\end{split}
\label{loc_func}
\end{equation}
corresponding, respectively, to the eigenvalues of the lower
hidden spectrum $\sigma_-=- 2t \cosh\eta$ and to the eigenvalues
of the upper hidden spectrum $\sigma_+= 2t \cosh\eta$. In Eqs.
(\ref{loc_func}), $A$ and $B$ are normalization constants and
$\eta \equiv 1/\xi$ is a parameter accounting for the localization
of the states. By rewriting Eq. (\ref{non_degenere}) for
$h=i\eta$, one has to solve:
\begin{equation}
(p-2)\cosh(\eta)-p  \coth{[\eta (L+1)]} \sinh{(\eta)}+2 \cos(k_0)
=0, \label{eta_legato}
\end{equation}
together with the condition that
$\sum_{x,y}\vert\psi_0(x,y)\vert^2=1$. In the thermodynamic limit,
$L \to \infty$, Eq. (\ref{eta_legato}) becomes:
$(p-2)\cosh(\eta)-p \sinh{(\eta)}+2 \cos(k_0)=0$, which is solved
by $\eta = \log\left[p - 1 + 2 \cos k_0 (p - 1 + \cos^2 k_0) + 2
\cos^2 k_0 \right]$, yielding, for the lower hidden spectrum:
$\sigma_-=-t \, \left[(p- 2)\cos(k_0) + p \sqrt{p - 1 +
\cos^2(k_0)}\right]/(p -1)$. The lowest energy level is obtained
for $\cos(k_0)=1$; one has
 \begin{equation}
E_0=-t\, \frac{(p- 2)+ p\,  \sqrt{p}}{p -1}.
\label{energia_ground_state}
\end{equation}
Solving the eigenvalue equation (\ref{eq1b}) for $E=E_0$, one
obtains the wavefunction of the localized ground-state:
\begin{equation}
\psi_{E_0}(y)=\sqrt{\frac{\sqrt{p}+2}{2\left(\sqrt{p}+1\right)}}\,
e^{- y/\xi}.
\end{equation}
$\xi=1/\log{(\sqrt{p}+1)}$ provides an estimate of the
ground-state localization. The ground state is localized along the
backbone and it decreases exponentially along the arms. When
$p=2$, one immediately recovers all the known results of the comb
graph \cite{burioni00,giusiano04}. For each value of $k_0$
($\cos(k_0)>0$) there is a solution of \eqref{eta_legato} with a
different energy in the interval $[E_0, -2t]$. In a finite
star-comb with $N_s(p L+1)$ sites there are $N_s$ solution of this
type and for $N_s \to \infty$ these solutions fill densely the
interval $[E_0, -2t ]$. Analogously, when $\cos(k_0)<0$ there are
$N_s$ solution of \eqref{eta_legato} corresponding to the spectral
region at high energy ($E=[2t, \vert E_0 \vert]$), where the other
hidden states appear.
\begin{figure}[t]
\includegraphics[scale=0.7]{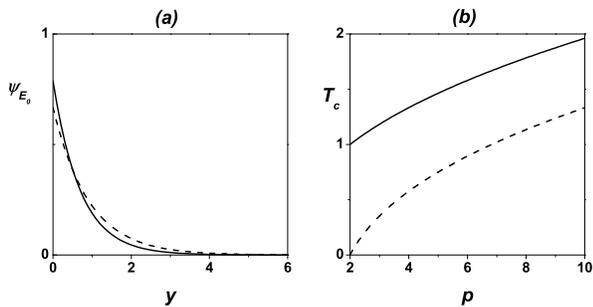}
\caption{\label{fig2_3} (a) The normalized single-particle
ground-state wavefunction for bosons hopping on a star and
star-comb graphs as a function of the distance $y$ from the
center. The number of arms is $p=10$. (b) Critical temperature
$T_c$ (in units of $k_B/E_J$) as a function of the number of arms
$p$. The solid lines are for the star-comb graph, whereas the
dashed lines are for the single star.} \vspace{8mm}
\end{figure}
In Fig. \ref{fig2_3}a we plot the ground-state wavefunction for
$p=10$ for a single star (dashed line) and for a star-comb (solid
line) as a function of the distance from the backbone. Figure
\ref{fig2_3}a evidences that adding stars enhances the
localization of the wavefunction around the center of each star.


The thermodynamic properties of non-interacting bosons hopping on
a star-comb graph evidence also in this case a topology induced
spatial BEC even if $d<2$ \cite{burioni00,burioni01}. To elucidate
this phenomenon, it is most convenient to introduce the
macrocanonical ensemble to determine the fugacity
$z=e^{\beta(\mu-E_0)}$ as a function of the temperature
\cite{burioni00}: the equation determining $z$ is given by
\begin{equation}
N_T=\sum_{E\in \sigma} \frac{d(E)}{z^{-1} e^{\beta (E-E_0)}-1}.
\label{filling1}
\end{equation}
In Eq. \eqref{filling1}, $d(E)$ is the degeneracy of each
single-particle eigenstate of the Hamiltonian (\ref{pure_hopping})
and $\beta=1/k_B T$. The sum in Eq. (\ref{filling1}) is over the
entire spectrum $\sigma$. For free bosons hopping on a star-comb
graph, one has
\begin{equation}
\label{filling}
\begin{split}
N_T &=N_{E_0}(L,T)+N_{\sigma_{-}}(L,T) +N_{\sigma_{+}}(L,T) 
\\
&+
\int_{E \in \sigma_0} dE \, \frac{N_s(p L+1) \rho(E)}{z^{-1}
e^{\beta (E-E_0)}-1 },
 \end{split}
\end{equation}
where $N_{E_0}(L,T)$, $N_{\sigma_{-}}(L,T)$ and
$N_{\sigma_{+}}(L,T)$ denote, respectively, the number of
particles at a certain temperature $T$ in the ground-state and in
the two regions $\sigma_{-}$ and $\sigma_{+}$ of the hidden
spectrum. $\rho(E)$, with $E\in \sigma_0$, is the energy density
of states of the linear chain, i.e.,
$\rho=1/(\pi\sqrt{4t^2-E^2})$, with $\sigma_0$ being the region of
the spectrum corresponding to delocalized states. The last term of
the right-hand side of Eq. \eqref{filling} is the number of bosons
in the delocalized (chain-like) states. The presence of the hidden
spectrum changes the behavior of the integral evaluated in the
interval $\{-2t,2t\}$, since it reduces it to the one describing
non-interacting bosons on a linear chain with an impurity in one
of the sites. As a result, letting $z \to 1$, the integral
converges even at finite temperatures making possible a spatial
BEC in $d<2$.


If one defines $T_c$ as the critical temperature at which BEC
occurs, then for any  $T<T_c$ the ground-state is macroscopically
filled. Since, at the critical temperature,
$N_{E_0}(T_c)=N_{\sigma_{\pm}}(T_c)=0 $, the equation allowing to
determine $T_c$ as a function of the parameters $f$ and $t$ reads:
\begin{equation}
\pi f=\int_{-2t}^{2t} \frac{dE}{\sqrt{{4t^2-E^2}} }
\frac{1}{e^{(E-E_0)/(k_{B}T_{c})}-1 }. \label{critical_T}
\end{equation}
Equation (\ref{critical_T}) can be solved numerically for any
value of $f$.  When $f \gg 1$, one may expand the exponential in
Eq. (\ref{critical_T}) to the first order in the inverse of the
critical temperature $T_c$. Upon inserting the value of the
ground-state energy (\ref{energia_ground_state}) in Eq.
(\ref{critical_T}), one easily finds that the critical temperature
$T_c$ is given by
\begin{equation}
 T_c \approx \frac{t f}{k_B} \frac{\sqrt{p}\,(2+\sqrt{p})}{1+\sqrt{p}}.
\label{t_c_f}
\end{equation}
Equation (\ref{t_c_f}) has been checked numerically and it is in
excellent agreement with the numerical solution of Eq.
(\ref{critical_T}) for $f \gg 1$, the error being of order $1/f$.
If one compares the critical temperature of the single star and of
the star-comb graphs, one immediately sees that, once the number
of arms is fixed, $T_c$  is always enhanced in the latter
realization. This is shown in Fig. \ref{fig2_3}b.

One may now use Eq. (\ref{t_c_f}) to determine the condensate
fraction as a function of the scaled temperature $T/T_c$. In the
thermodynamic limit, the number of particles in the delocalized
states is given by
\begin{equation}
\begin{split}
N_{\sigma_{0}}&= \lim_{L \to \infty} N_s(pL+1) \int_{E \in
\sigma_0} \rho(E) \frac{dE}{e^{\beta (E-E_0)}-1 }
\\
&\approx N_T \frac{T}{T_c}.
\end{split}
 \label{n_B}
\end{equation}
In the last equation the exponential has been expanded to the
first order in $\beta$: this approximation holds for $f \gg 1$ and
it is in very good agreement with the numerical evaluation of the
integral in Eq. (\ref{n_B}) also in a large neighborhood below
$T_c$. From Eqs. (\ref{filling}) and (\ref{n_B}) one gets the
number of particles in the localized states $N_0=N_{E_0}+
N_{\sigma_{-}}$: the fraction of condensate, for $T<T_c$, is then
given by
\begin{equation}
\frac{N_0}{N_T} \approx 1 - \frac{T}{T_c}. \label{n_0}
\end{equation}
For $f$ ranging from $10^3$ to $10^9$, the results provided by Eq.
(\ref{n_0}) differ from those obtained by the numerical evaluation
of $N_0$ from Eq. (\ref{filling}) by less than $1 \%$. Equation
(\ref{n_0}) clearly shows that the condensate has dimension $1$
just as cigar-shaped one-dimensional atomic Bose condensates
\cite{ketterle96_gorlitz01}.


Due to the topology induced spatial condensation on the backbone,
one should expect an inhomogeneous distribution of the bosons
along the arms of the network. The average number of bosons $N_B$
at a site $(x,y,z)$ depends - due to the symmetry of the graph -
only on the distance $y$ from the backbone. It is not difficult to
show (see also Ref. \cite{giusiano04}) that, away from the
backbone ($y \gg 1$) and once the filling is fixed, $N_B$ depends
only on the scaled temperature $T/T_c$ and it is given by
\begin{equation}
\label{rapporto_sempl} \frac{N_B (y;T/T_c)}{f} \approx
\frac{T}{T_c}.
\end{equation}
Topology induced spatial BEC in a system of non-interacting bosons
hopping on a star-comb graph predicts then a rather sharp decrease
of the number of bosons at sites located away from the backbone.
The linear dependence is consistent with the observation that, in
this system, the condensate has dimension 1.


In this paper we have investigated in detail another situation
where the role played by the network's topology is crucial in
determining the thermodynamic properties of the system. We focused
on the properties of free bosons hopping on a star-comb network,
finding an enhancement  - with respect to other inhomogeneous
structures previously investigated
\cite{burioni00,burioni01,giusiano04,brunelli04} - of the critical
temperature at which the particles condense.


\begin{thebibliography}{99}


\bibitem{burioni00}
R. Burioni, D. Cassi, I. Meccoli, M. Rasetti, S. Regina, P.
Sodano, and A. Vezzani, Europhys. Lett. {\bf 52}, 251 (2000).

\bibitem{burioni01}
R. Burioni, D. Cassi, M. Rasetti, P. Sodano, and A. Vezzani, J.
Phys. B {\bf 34}, 4697 (2001).

\bibitem{giusiano04}
G. Giusiano, F. P. Mancini, P. Sodano, and A. Trombettoni, Int. J.
Mod. Phys. B  {\bf 18}, 691 (2004).

\bibitem{brunelli04}
I. Brunelli, G. Giusiano, F. P. Mancini, P. Sodano, and A.
Trombettoni, J. Phys. {\bf B 37}, S275 (2004).

\bibitem{ketterle96_gorlitz01} W. Ketterle and N. J. van Druten,
Phys. Rev. A {\bf 54}, 656 (1996); A. G\"orlitz {\em et al.},
Phys. Rev. Lett. {\bf 87}, 130402 (2001).



\end{thebibliography}
\end{document}